\documentclass[showpacs,rotating,preprintnumbers,amsmath,amssymb]{elsart}
\usepackage{rotating}
\usepackage{graphicx}
\usepackage{dcolumn,color}
\usepackage{bm}
\usepackage{epsfig}
\begin{document}
\bibliographystyle{try}

\newcounter{univ_counter}
\setcounter{univ_counter} {0}
\addtocounter{univ_counter} {1}
\edef\HISKP{$^{\arabic{univ_counter}}$ } \addtocounter{univ_counter}{1}
\edef\GATCHINA{$^{\arabic{univ_counter}}$ } \addtocounter{univ_counter}{1}
\edef\KVI{$^{\arabic{univ_counter}}$ } \addtocounter{univ_counter}{1}
\edef\PI{$^{\arabic{univ_counter}}$ } \addtocounter{univ_counter}{1}
\edef\GIESSEN{$^{\arabic{univ_counter}}$ } \addtocounter{univ_counter}{1}
\edef\BASEL{$^{\arabic{univ_counter}}$ } \addtocounter{univ_counter}{1}
\edef\FSU{$^{\arabic{univ_counter}}$ } \addtocounter{univ_counter}{1}

\begin{frontmatter}

\title{Photoproduction of meson pairs: First measurement of the polarization observable \boldmath $I^{s}$\unboldmath }

\collab{The CBELSA/TAPS Collaboration}

\author[HISKP]{E.~Gutz}$^,$\footnote[1]{Corresponding authors. \textit{E-mail addresses:}\\ {\tt gutz@hiskp.uni-bonn.de} (E. Gutz)\\ {\tt thoma@hiskp.uni-bonn.de} (U. Thoma)},
\author[HISKP]{V.~Sokhoyan},
\author[HISKP]{H.~van~Pee},
\author[HISKP,GATCHINA]{A.V.~Anisovich},
\author[KVI] {J.C.S.~Bacelar},
\author[PI]{B.~Bantes},
\author[HISKP]{O.~Bartholomy}, 
\author[HISKP,GATCHINA]{D.~Bayadilov},
\author[HISKP]{R.~Beck},
\author[GATCHINA]{Yu.~Beloglazov},
\author[KVI]{R.~Castelijns},
\author[HISKP,FSU]{V.~Crede},
\author[PI]{H.~Dutz},
\author[PI]{D.~Elsner},
\author[PI]{R.~Ewald},
\author[PI]{F.~Frommberger},
\author[HISKP]{M.~Fuchs},
\author[HISKP]{Ch.~Funke},
\author[GIESSEN]{R.~Gregor},
\author[GATCHINA]{A.~Gridnev},
\author[PI]{W.~Hillert},
\author[HISKP]{Ph. Hoffmeister},
\author[HISKP]{I.~Horn},
\author[BASEL]{I.~J\"agle},
\author[HISKP]{J.~Junkersfeld},
\author[HISKP]{H.~Kalinowsky},
\author[PI]{S.~Kammer},
\author[PI]{V.~Kleber}$^,$\footnote[2]{{\em Present address:} German Research School for Simulation Sciences, J\"{u}lich, Germany},
\author[PI]{Frank~Klein},
\author[PI]{Friedrich~Klein},
\author[HISKP]{E.~Klempt},
\author[GIESSEN,BASEL]{M.~Kotulla},
\author[BASEL]{B.~Krusche},
\author[HISKP]{M.~Lang},
\author[KVI]{H.~L\"ohner},
\author[GATCHINA]{I.~Lopatin},
\author[GIESSEN]{S.~Lugert},
\author[PI]{D.~Menze},
\author[BASEL]{T.~Mertens},
\author[KVI]{ J.G.~Messchendorp},
\author[GIESSEN]{V.~Metag},
\author[GIESSEN]{M.~Nanova},
\author[HISKP,GATCHINA]{V.~Nikonov},
\author[GATCHINA]{D.~Novinski},
\author[GIESSEN]{R.~Novotny},
\author[PI]{M.~Ostrick}$^,$\footnote[3]{{\em Present address:} Institut f\"{u}r Kernphysik, Universit\"{a}t Mainz, Germany},
\author[GIESSEN]{L.~Pant}$^,$\footnote[4]{{\em On leave from:} Nucl. Phys. Division, BARC, Mumbai, India},
\author[GIESSEN]{M.~Pfeiffer},
\author[HISKP]{D.~Piontek},
\author[FSU]{W.~Roberts},
\author[GIESSEN]{A.~Roy}$^,$\footnote[5]{{\em On leave from:} Department of Physics, IIT, Mumbai, India},
\author[HISKP,GATCHINA]{A.~Sarantsev},
\author[GIESSEN]{S.~Schadmand}$^,$\footnote[6]{{\em Present address:} Institut f\"{u}r Kernphysik and J\"{u}lich Center for Hadron Physics, Forschungszentrum J\"{u}lich, Germany},
\author[HISKP]{Ch.~Schmidt},
\author[PI]{H.~Schmieden},
\author[PI]{B.~Schoch},
\author[KVI]{S.~Shende},
\author[PI]{A.~S\"ule},
\author[GATCHINA]{V.~Sumachev},
\author[HISKP]{T.~Szczepanek},
\author[HISKP]{A.~Thiel},
\author[HISKP]{U.~Thoma}$^,$\footnotemark[1],
\author[GIESSEN]{D.~Trnka},
\author[GIESSEN]{R.~Varma}$^,$\footnotemark[5],
\author[PI]{D.~Walther},
\author[HISKP]{Ch.~Weinheimer}$^,$\footnote[7]{{\em Present address:} Institut f\"{u}r Kernphysik, Universit\"{a}t M\"{u}nster, Germany},
\author[HISKP]{Ch.~Wendel}

\address[HISKP]{Helmholtz-Institut f\"ur Strahlen- und Kernphysik, Universit\"at Bonn, Germany}
\address[GATCHINA]{Petersburg Nuclear Physics Institute, Gatchina, Russia}
\address[KVI]{Kernfysisch Verseller Instituut, Groningen, The Netherlands}
\address[PI]{Physikalisches Institut, Universit\"at Bonn, Germany}
\address[GIESSEN]{II. Physikalisches Institut, Universit\"at Giessen, Germany}
\address[BASEL]{Physikalisches Institut, Universit\"at Basel, Switzerland}
\address[FSU]{Department of Physics, Florida State University, Tallahassee, FL 32306, USA}

\date{\today}
\newpage
\begin{abstract}
The polarization observable $I^{s}$, a feature exclusive to the acoplanar kinematics of multi-meson final states produced via linearly polarized photons, has been measured for the first time. Results for the reaction $\vec{\gamma}\mathrm{p}\rightarrow \mathrm{p} \pi^{0} \eta$ are presented for incoming photon energies between 970\,MeV and 1650\,MeV along with the beam asymmetry $I^{c}$. The comparably large asymmetries demonstrate a high sensitivity of $I^{s}$ to the dynamics of the reaction. The sensitivity of these new polarization observables to the contributing partial waves is demonstrated by fits using the Bonn-Gatchina partial wave analysis. \vspace{5mm}   
\end{abstract}
\end{frontmatter}

\begin{figure}[pt]
\includegraphics[width=\textwidth]{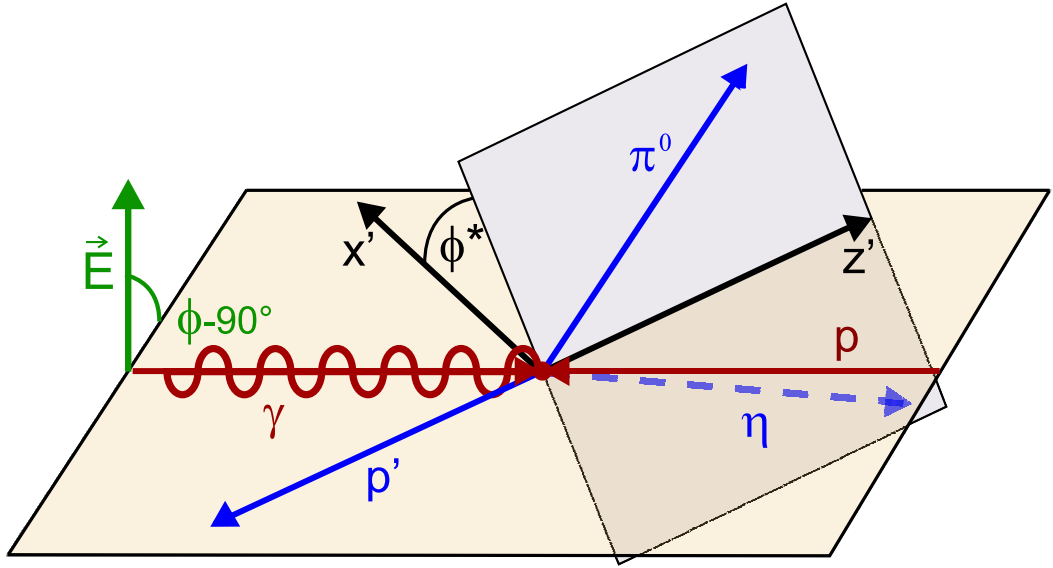} \\
\caption{Angle definitions in the center-of-momentum frame.
$\phi^{*}$ is the angle between the reaction plane defined by the
incoming photon and recoiling particle $p'$ and the
decay plane of two final state particles.}
\label{fig:phidef}
\end{figure}

\begin{figure}[pt]
\begin{center}
\includegraphics[width=.75\textwidth]{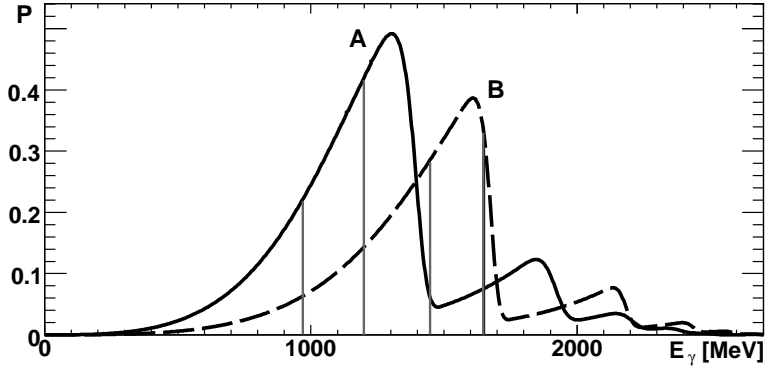} \\
\end{center}
\caption{Degree of linear polarization for the two settings. The
largest polarizations were 49.2\% at $E_\gamma = 1300$\,MeV (A) and
38.7\% at 1600\,MeV (B), respectively (see \cite{Elsner-EPJA2} for
details). Vertical lines indicate the chosen energy ranges.}
\label{fig:pol}
\end{figure}

\begin{figure}[pt]
\begin{center}
\includegraphics[width=.75\textwidth]{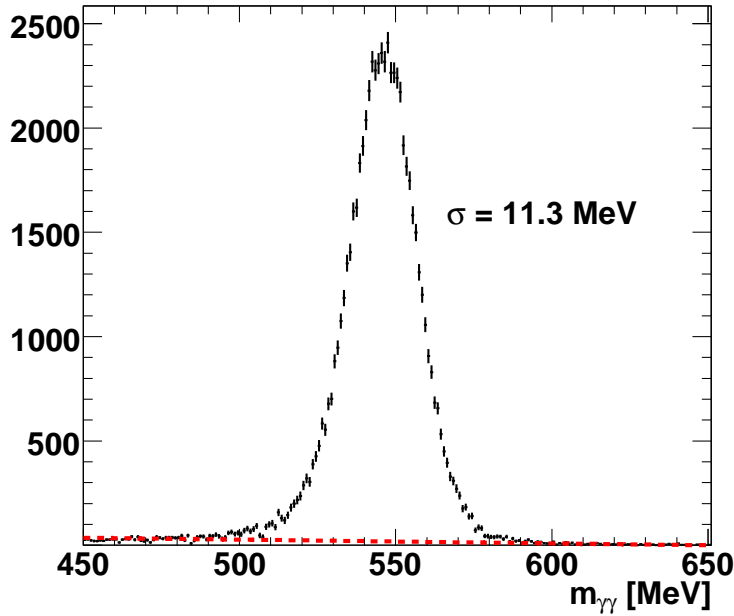} \\
\end{center}
\caption{$\gamma\gamma$ invariant mass distribution after cuts on
the confidence levels of the $\gamma \mathrm{p} \rightarrow
\mathrm{p} \pi^{0} \gamma \gamma$ ($\mathrm{CL} > 8\%$ ) and $\gamma
\mathrm{p} \rightarrow \mathrm{p} \pi^{0} \pi^{0}$ ($\mathrm{CL} <
1\%$ ) fits respectively. This yields a total number of 68514 events
including the linear background (red line). An additional cut on the
$\gamma \mathrm{p} \rightarrow \mathrm{p} \pi^{0} \eta$ fit
($\mathrm{CL} > 6\%$) rejects 3083 events, retaining 624 background
events ($\approx 1\%$).}
\label{fig:eta}
\end{figure}

\begin{figure}[pt]
\begin{center}
\includegraphics[width=.75\textwidth]{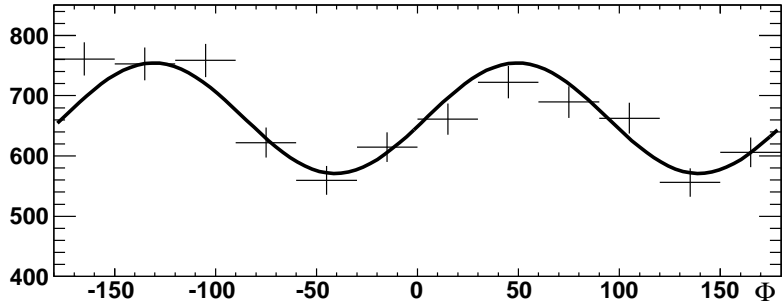} \\
\end{center}
\caption{Example of a measured $\phi$-distribution. Shown is the
$\phi$-distribution of the final state proton in the region
$60^{\circ} \le \phi^{*}\le 120^{\circ}$ for events in the energy
range $W = 1834 \pm 64$\,MeV (y-axis with suppressed-zero scale).}
\label{fig:phi}
\end{figure}

\begin{figure}[pt]
\includegraphics[width=\textwidth]{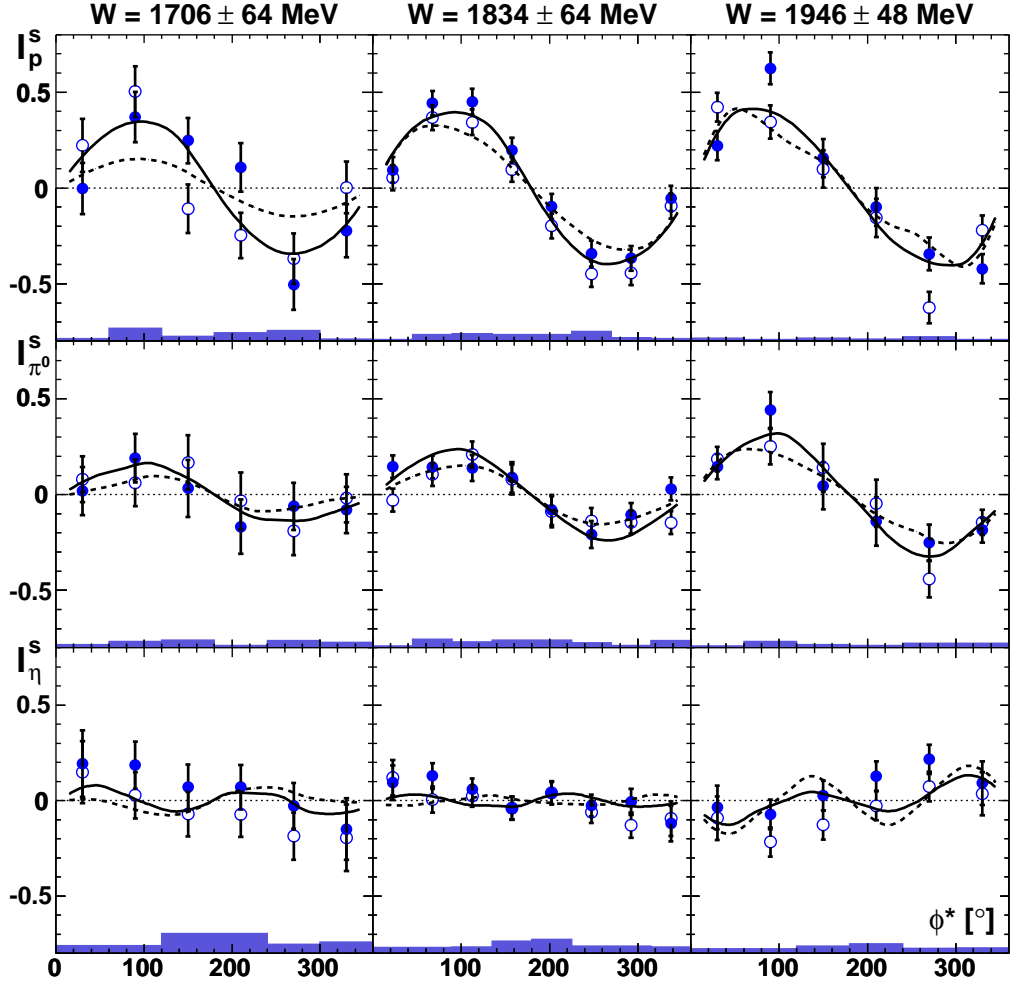}
\caption{Measured beam asymmetries $I^{s}$ in the reaction
$\vec{\gamma} \mathrm{p}\rightarrow\mathrm{p}\pi^{0}\eta$. Left to
right: CMS energy ranges $1706\pm 64$\,MeV, $1834\pm 64$\,MeV,
$1946\pm 48$\,MeV. Top to bottom: Beam asymmetries obtained treating
the proton (top row), $\pi^{0}$ (center row) and $\eta$ (bottom row)
as recoiling particle. Filled symbols: $I^{s}(\phi^{*})$, open
symbols: $-I^{s}(2\pi - \phi^{*})$. Solid curve: Full BnGa-PWA fit, dashed curve: BnGa-PWA
fit excluding $3/2^{-}$-wave. Histograms below: Estimate of
systematic errors due to acceptance and efficiency.}
\label{fig:is}
\end{figure}

\begin{figure}
\includegraphics[width=\textwidth]{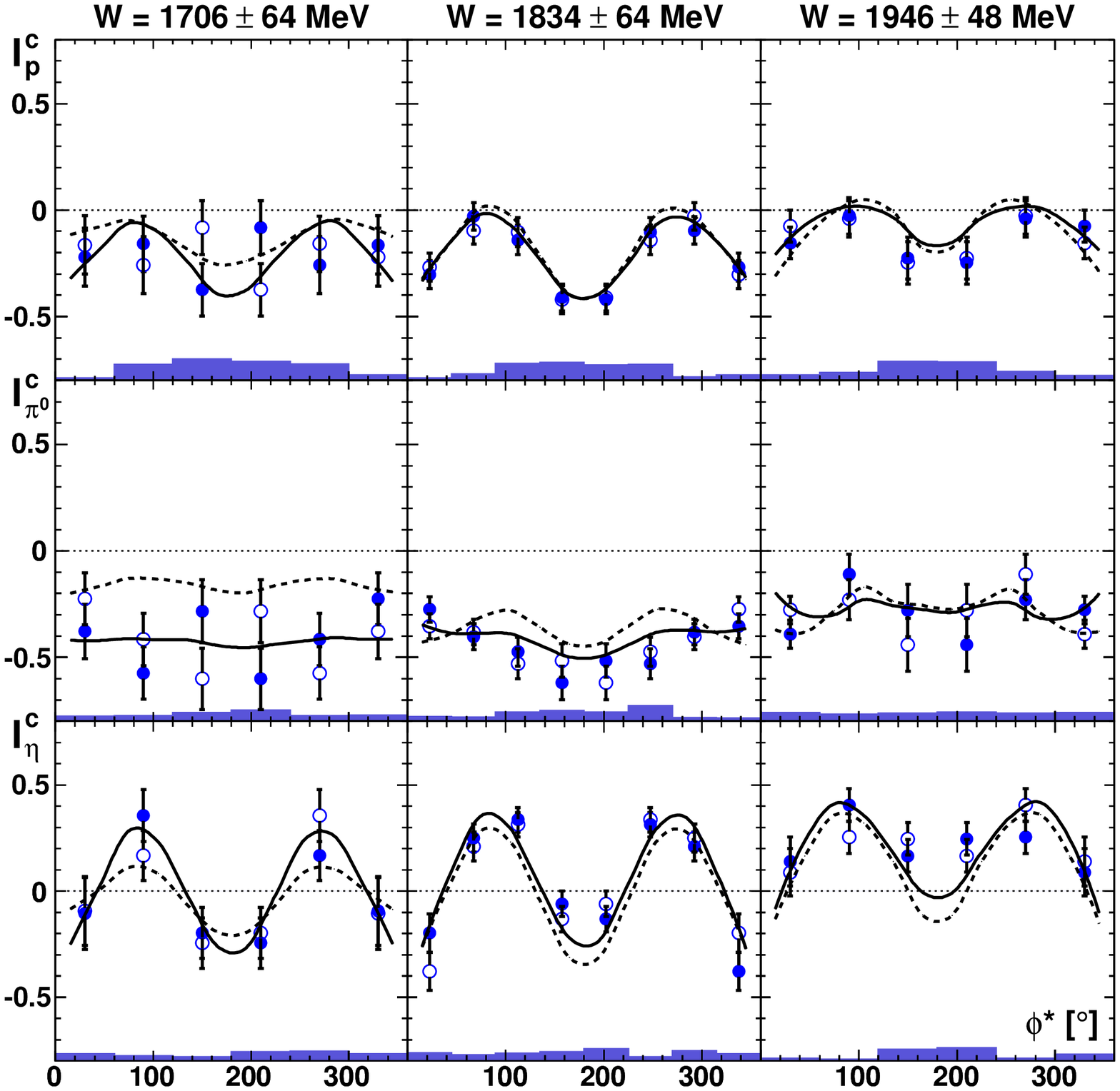}
\caption{Measured beam asymmetries $I^{c}$ in the
reaction $\vec{\gamma}\mathrm{p}\rightarrow\mathrm{p}\pi^{0}\eta$.
Notation as Fig.~\ref{fig:is}, except filled symbols:
$I^{c}(\phi^{*})$, open symbols: $I^{c}(2\pi - \phi^{*})$.}
\label{fig:ic}
\end{figure}

Baryons manifest the non-Abelian nature
of the strong interaction. Thus, study of baryon excited states and
production processes can provide insight into the dynamics and
degrees of freedom relevant for non-perturbative quantum
chromodynamics (QCD). At present, much of our limited understanding
of these excited states comes from symmetric quark models
\cite{CapRob-PPNP,Loering-EPJA}. These models predict a number of
states with masses above 1.8\,GeV that have not been observed in the
$\pi N$ channel \cite{CapRob-PRD}, the so-called {\it missing
resonances}. Photoproduction of multi-meson final states avoids $\pi
N$ in the initial and the final state and gives the opportunity to
probe the sequential decays of such high-lying resonances. Especially in the regime of excited $\Delta$ states the
$\Delta \eta$ final state is particularly attractive due to its
isospin selectivity.  Accordingly, the study of the photoproduction of
multi-meson final states and in particular the reaction
\begin{equation}
\label{reaction}
\gamma \mathrm{p} \rightarrow \mathrm{p} \pi^{0} \eta
\end{equation}
has gained in importance over the past years, both from the
experimental  side with the measurement of unpolarized total and
differential cross sections
\cite{Nakabayashi:2006ut,Ajaka:2008zz,Horn-PRL,Horn-EPJA,Kashevarov-EPJA}
and the beam asymmetry $\Sigma$ \cite{Ajaka:2008zz,Gutz-EPJA}, as well as
from the theoretical side. In the low-energy region, there have been attempts to
treat the $\Delta(1700)D_{33}$ as resonance that is dynamically generated from $\Delta$-$\eta$
interactions \cite{Doering-PRC}, as well as attempts to understand
the rapidly rising cross section \cite{Fix-EPJA} by formation of
intermediate resonances. In the Bonn Gatchina partial wave analysis (BnGa-PWA),
described in \cite{Anisovich-EPJA1,Anisovich-EPJA2}, evidence was
reported for the $\Delta(1920)$, an established (three-star) resonance
in the $J^P=3/2^+$-wave and a not-well-known (one-star)
resonance $\Delta(1940)$ with spin and parity $J^P=3/2^-$ \cite{Horn-PRL,Horn-EPJA}.
The two resonances seem to form a further parity doublet, possibly
indicating a restoration of chiral symmetry at high baryon
excitation masses \cite{Glozman:1999tk}. The mass of the $J^P=3/2^-$-state 
indicates a mild conflict with quark models \cite{CapRob-PPNP,Loering-EPJA} and
is consistent with models describing QCD in terms of a dual gravitational
theory, AdS/QCD \cite{Forkel:2008un,deTeramond:2009xk}.\\
Two-meson photoproduction is not - like two-body reactions -
restricted to a single plane as seen in Fig.~\ref{fig:phidef}; two
planes, a reaction and a decay plane enclosing an angle $\phi^{*}$,
occur. In contrast to single-meson production, here polarization asymmetries can also occur if e.g. only the target is
longitudinally polarized or if only the beam is circularly polarized. The first measurements of the latter asymmetries in double-pion photoproduction \cite{Strauch-PRL,Krambrich-PRL} have demonstrated their significant model sensitivity and revealed serious deficiencies of most available models. 
For linearly polarized photons impinging on an unpolarized target
two polarization observables $I^{s}$ and $I^{c}$ occur, for which so far no data has been published in any channel. The
latter corresponds to the polarization observable $\Sigma$ if the
dependence on the angle $\phi^*$ is integrated out. The cross section
is written as 
\begin{equation}
\label{xsect}
\frac{\mathrm{d}\sigma}{\mathrm{d}\Omega} =
\left(\frac{\mathrm{d}\sigma} {\mathrm{d}\Omega}\right)_{0}\{1 +
\delta_{l}[I^{s}\sin(2\phi)+I^{c}\cos(2\phi)]\}\,,
\end{equation}
\cite{RobOed-PRC} where $\left(\frac{\mathrm{d}\sigma}{\mathrm{d}\Omega}\right)_{0}$ 
is the unpolarized cross section, $\delta_{l}$ is the
degree of linear photon polarization, and $\phi$ the azimuthal
angle of the reaction plane with respect to the normal on the polarization plane. 
Since polarization observables are very sensitive to
interference effects in the amplitudes, they are
expected to significantly constrain reaction models, and hence make the extraction of resonance parameters much
more precise than unpolarized data alone would allow. Furthermore,
observables such as $I^{s}$ ($I^{c}$) can be expressed as the
imaginary (real) part of a linear combination of bilinears formed from the helicity or
transversity amplitudes that describe the process. They are therefore
not only particularly sensitive to interference effects, but also to
the relative phases of the amplitudes.\\
The data were obtained using the tagged photon beam of the ELectron Stretcher Accelerator (ELSA)~\cite{Hillert-EPJA} and the CBELSA/TAPS detector.
The experimental setup consists of an arrangement of two
electromagnetic calorimeters, the Crystal Barrel detector
\cite{Aker-NIM} comprising 1290 CsI(Tl) crystals and the TAPS
detector \cite{Nowotny-IEEE,Gabler-NIM} in a forward wall setup
consisting of 528 BaF$_2$ modules in combination with plastic
scintillators for charge information. Together these
ca\-lo\-ri\-me\-ters cover the polar angular range from $5^\circ$ to
$168^\circ$  and the full azimuthal range. For further charged
particle identification a three layer scintillating fiber detector
\cite{Suft-NIM} surrounds the 5\,cm long liquid hydrogen target
\cite{Kopf-PhD}.\\
The linearly polarized photons are produced via coherent
bremsstrahlung of the initial 3.2\,GeV electron beam off a diamond
radiator. Electrons undergoing the bremsstrahlung process are then
momentum analyzed using a tagging spectrometer consisting of a
dipole magnet and a scintillator based detection system. For further
details on the experimental setup, see \cite{Elsner-EPJA1}.\\
For this analysis, two datasets were considered. Fig.~\ref{fig:pol}
shows the degree of polarization as a function of the incident
photon energy for two diamond radiator orientations. The systematic error 
of the polarization was determined to be $\Delta\mathrm{P}\le 0.02$ \cite{Elsner-EPJA2}.
The two datasets were subdivided into three energy ranges, $W =
1706\pm 64$\,MeV, $1834\pm 64$\,MeV, and $1946\pm 48$\,MeV
respectively, as indicated by the vertical lines in Fig.~\ref{fig:pol}.
 To guarantee a sufficiently high degree of polarization, the low energy range consists solely of data taken
with the polarization setting A, the high energy range of data taken
with setting B. For the intermediate energy range, both datasets
were combined. To select the reaction (\ref{reaction}), events with five
distinct hits in the calorimeters were considered in further analysis. Events were retained if at least
one combination of four out of the five clusters was consistent with
a $\pi^0$ and an $\eta$ in the final state as determined by a
4$\sigma$ cut on the corresponding two-particle invariant mass
distributions. To avoid possible systematic effects due to scintillator inefficiencies, charge information was not used to identify the proton. Instead, the direction of the fifth particle had to agree with the missing momentum of the supposed two-meson system; the angular difference had to be smaller than $10^{\circ}$ in $\phi$ and, depending on the
angular resolution in the polar angle of the calorimeters, $5^{\circ}$ in 
$\theta$ for TAPS and $15^{\circ}$ for the Crystal Barrel, respectively. Additionally the missing mass
needed to be consistent with the proton mass within 4$\sigma$.
After applying the preselection, the data was subjected to a kinematic fit \cite{Pee-EPJA} imposing energy and momentum conservation, assuming that the interaction took place in the target center. Only events that exceeded, according to the respective distributions, a probability (CL) of 8\% for the $\gamma \mathrm{p} \rightarrow \mathrm{p} \pi^{0}\gamma \gamma$ two-constraint hypothesis and of 6\% for the $\gamma \mathrm{p} \rightarrow \mathrm{p} \pi^{0} \eta$ three-constraint hypothesis, respectively, were retained. The proton direction resulting from the fit had to agree with the direction of the proton determined as stated above within 20$^{\circ}$. In addition, events compatible with $\mathrm{CL} > 1\%$ for the $\gamma \mathrm{p} \rightarrow\mathrm{p} \pi^{0} \pi^{0}$ hypothesis were rejected. The final event sample contains a total of 65431 events from reaction (\ref{reaction}) with a maximum background contamination of 1\%
(Fig.~\ref{fig:eta}).
To extract the polarization observables defined in Eq.~(\ref{xsect}), the $\phi$ distribution of the final state
particles was fit with the expression
\begin{equation}
f(\phi) = A + P\,[B\,\sin(2\phi) + C\,\cos(2\phi)],
\end{equation}
with $P$ being the polarization determined for each event
individually and later averaged for each fitted bin. Fig.~\ref{fig:phi} shows an example of an according distribution. The
effect of both beam asymmetries is clearly visible in the distinct
superposition of a $\cos(2\phi)$- ($I^{c}$) and a $\sin(2\phi)$-modulation ($I^{s}$).
From the fits to the according $\phi$-distributions $I^{s}$ and
$I^{c}$ have been extracted, as shown in Figs.~\ref{fig:is},~\ref{fig:ic}.\\
When investigating asymmetries,  the detection efficiency is usually
considered not to have an influence on the result. In the quotients
$B/A$ or $C/A$ this drops out as long as the bins in the 5-dimensional
phase space can be considered reasonably small compared to the
variation of efficiency. If on the other hand the  5-dimensional
phase space is not completely covered, which is true for most of the
experiments, the given distributions represent only the
polarization observable within the covered phase space. The
acceptance for the CBELSA/TAPS experiment determined from MC
simulations vanishes for forward protons leaving TAPS
through the forward hole, and for protons going backward in the
center-of-mass system, having very low laboratory momenta. 
To study these effects on the shown distributions, different
MC datasets have been produced and analyzed. 
First of all a phase space MC dataset has been produced and was analyzed using the same
analysis chain as for the data. A 2-dimensional acceptance and efficiency correction as function of the variables $\phi$ and $\phi^{*}$ 
has been determined. 
In addition, since effects due to the contributing physics amplitudes have to be considered, the result of the
PWA discussed below has been used to study the acceptance and efficiency. The systematic error shown in 
Figs.~\ref{fig:is} and \ref{fig:ic} reflects the maximal effect determined 
by these methods. Given the statistical uncertainties
of the data points the effects due to the acceptance and efficiency correction are
small. \\
Symmetry properties allow for a further cross check of the data. $I^{s}$ has to vanish for coplanar kinematics ($I^{s}(\phi^{*} = 0)= I^{s}(\phi^{*} = \pi) = I^{s}(\phi^{*} = 2\pi) =0$) and the transition $\phi^{*}\rightarrow2\pi-\phi^{*}$ is equivalent to a mirror operation with respect to the reaction plane. In the case of linear polarization this leads to the transition $\phi \rightarrow2\pi - \phi$ and because $\sin(2\cdot(2\pi-\phi)) = -\sin(2\phi)$ to $I^{s} \rightarrow -I^{s} $ (see Eq.~\ref{xsect}). These symmetry properties are clearly visible in the data with deviations consistent with statistics, which again shows the comparably small systematic uncertainties.\\ 
The sensitivity of the data to partial wave contributions is tested within the BnGa multi-channel partial wave analysis. The BnGa-PWA fits include a large number of reactions; a survey of the presently used datasets can be found elsewhere. Included in this fit were data on the reaction $\gamma p\to p\pi^0\eta$ but without information on $I^{s}$ and $I^{c}$. The fit \cite{Horn-PRL,Horn-EPJA} had claimed evidence for contributions from negative- and positive-parity $\Delta$ resonances with spin $J=3/2$, the $\Delta(1700)$ and the poorly established $\Delta(1940)$  resonances with $J^P=3/2^-$, and the established $\Delta(1600)$ and $\Delta(1920)$ resonances with $J^P=3/2^+$. The result of a new fit including $I^{s}$ and $I^{c}$ is shown in Figs.~\ref{fig:is} and ~\ref{fig:ic} as solid curves. Removing the couplings of the $3/2^{+}$-wave to p$\pi^{0}\eta$ (which provides a small fraction of the total cross section only) results in a fit to $I^{s}$ and $I^{c}$ which is still acceptable; larger discrepancies are only observed in differential cross sections. However, removing the $3/2^{-}$-wave which includes the above mentioned resonances $\Delta(1700)$ and $\Delta(1940)$ leads to noticeable discrepancies in the fits, shown as dashed curves in Figs.~\ref{fig:is} and ~\ref{fig:ic}.\\
In addition to these fits within the BnGa-PWA, which demonstrate the sensitivity of $I^{s}$ and $I^{c}$ to the contributing partial waves, a preliminary comparison of the data with predictions using the chiral unitarity framework of \cite{Doering-PRC} shows a significant relation between these new polarization observables and the production dynamics (\ref{reaction}) \cite{Valencia}. Furthermore, discrepancies between these predictions and the data at higher energies point towards the need for additional contributions to be included in the model. These observations underline the importance of polarization observables in general and demonstrates the significance of $I^{s}$ and $I^{c}$ as new polarization observables in particular.\\

We thank the technical staff of ELSA and the par\-ti\-ci\-pating institutions for their invaluable contributions to the success of the experiment. We acknowledge financial support from the \textit{Deutsche Forschungsgemeinschaft} (SFB/TR16) and \textit{Schweizerischer Nationalfonds}.

\end{document}